\documentclass[twocolumn,aps,prb,showpacs,groupedaddress,amsfonts,amssymb,amsmath]{revtex4-2}
\usepackage{graphicx}
\usepackage{epstopdf}
\usepackage[colorlinks=true, pdfstartview=FitV, linkcolor=blue, citecolor=blue, urlcolor=blue]{hyperref}
\begin{document}
\title{Band alignment effect in the topological photonic alloy}
%\title{Topological Photonic Alloy Based on Perfect Electric Conductor and Magnetized Yttrium Iron Garnet}

\author{Tiantao Qu,$^{1}$ Mudi Wang,$^{2,*}$ Jun Chen,$^{1,4,\dagger}$ and Lei Zhang$^{3,4\ddagger}$}
\affiliation{$^1$State Key Laboratory of Quantum Optics and Quantum Optics Devices, Institute of Theoretical Physics, Shanxi University, Taiyuan 030006, China\\
	$^2$Department of Physics, The Hong Kong University of Science and Technology, Clear Water Bay, Hong Kong 999077, China\\
	$^3$State Key Laboratory of Quantum Optics and Quantum Optics Devices, Institute of Laser Spectroscopy, Shanxi University, Taiyuan 030006, China\\
	$^4$Collaborative Innovation Center of Extreme Optics, Shanxi University, Taiyuan 030006, China
}

\begin{abstract}
Recently, a photonic alloy with non-trivial topological properties has been proposed, based on the random mixing of Yttrium Iron Garnet (YIG) and magnetized YIG rods. When the doping concentration of magnetized YIG rods is less than one, a chiral edge state (CES) of the topological photonic alloy appears in the frequency range of the non-trivial topological gap of the magnetized YIG crystal. In this work, we surprisingly find that by randomly mixing the Perfect Electric Conductor (PEC) and magnetized YIG rods in a square lattice, the photonic alloy system with appropriate doping concentrations can present CES in special frequency intervals even when both components support the propagation of bulk states. Analyzing the band structure of two components, we noticed a shift between the first trivial bandgap for PEC and the first topological bandgap for magnetized YIG. When calculating the transmission spectrum of the photonic alloy, we discovered that the frequency range for the topological gap gradually opens from the lower limit frequency of the bandgap for PEC to the bandgap for the magnetized YIG rods. The topological gap opening occurs as the doping concentration of magnetized YIG rods increases, creating an effective band alignment effect. Moreover, the topological gap for the photonic alloy is confirmed by calculating the reflection phase winding with the scattering method. Lastly, the gradual appearance of the CES is identified by applying Fourier transformation to real-space electromagnetic fields. Our work broadens the possibilities for flexible topological gap engineering in the photonic alloy system.
\end{abstract}
\maketitle

\section{Introduction}

Over the past few decades, research on topological phenomena has grown exponentially. The most common approach involves the breaking or preservation of time reversal  and spatial inversion symmetry, enabling the realization of various topological effect \cite{Hasan1,Zhang2,Ozawa5,Lu6,Kim7,Xue2022}, such as quantum Hall effect \cite{Haldane48,Jotzu2014,Zhaoju17,He18}, quantum spin Hall effect \cite{Kane3,Zhang4,zhang2006}, quantum valley Hall effect \cite{Xiao2007,Rycerz2007,Hui2014,Mak2014}, etc. Photonic systems, due to their experimental feasibility, hold a crucial position in the discovery and experimental verification of various topological states \cite{Haldane8,Wang9,Skirlo45,Skirlo46,Hu2015,Barik11,Shalaev12,Yang13,Wang14,Ma_2016,Dong2017,Lan2023,Li2023,Jianfei2023,Zhang2024}.

In recent years, methods like topological quantum chemistry and symmetry indicators have facilitated rapid classification of topological states  \cite{Bradlyn2017,Liu2021,Benalcazar2019,dePaz2019}. This has led to extensive studies on the topological prevalence of periodically structured crystals \cite{Po2017,ghorashi2023prevalence}. However, as research evolves, there is an increasing focus on non-periodic systems, such as amorphous \cite{Mitchell25,Zhe26,Agarwala27,Yang28,Wang29,Ivaki30,Agarwala31,Zhou37,Bing38,xiaoyu}, Anderson insulating \cite{Stutzer34,Liu35,Li39,Groth40,Meier41,Cui2022,yanxia}, quasicrystalline systems \cite{Kraus32,Bandres33,Huang2018,Peng2022,Xu2023}, etc. Researchers employ real-space topological invariants such as Bott index, local Chern number, local Chern marker and scattering methods to characterize the topological properties of these non-periodic systems \cite{Huang2018,Zhou37,Liu35,Cui2022,Dixon2023,Ornellas2022}.

Mixing different types of photonic crystals can form a disordered photonic system, similar to semiconductor alloys in photonics \cite{PC_alloy1}. This system produces photonic localization, which can be applied to random lasers \cite{PC_alloy2,PC_alloy3}. Recently, a novel type of disordered topological photonic system called topological photonic alloy has been proposed by us \cite{TPA}. In a substitutional topological photonic alloy, a low-threshold topological gap is created by doping nonmagnetized Yttrium Iron Garnet (YIG) rods with magnetized YIG rods. This gap falls entirely within the bandgap frequency range of magnetized YIG crystals. The limitations imposed by the frequency range could potentially hinder the widespread application of topological photonic alloys. A nature question will arise: Is it possible for the frequency range of the topological alloy's bandgap to be outside the bandgap frequency range of the two crystals that compose it?

\begin{figure*}
	\centering
	\includegraphics[width=0.95\textwidth]{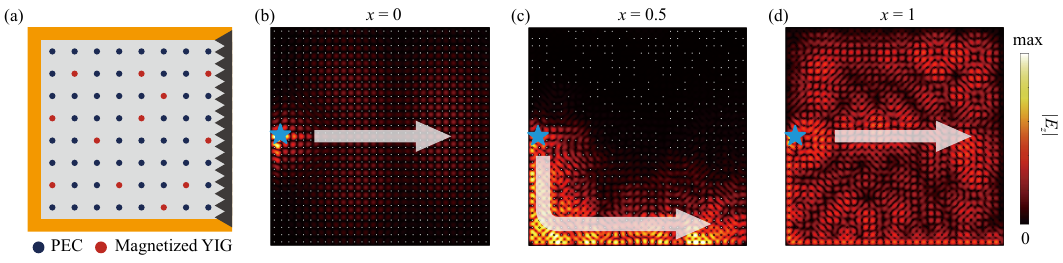}
	\caption{(a)  A schematic diagram of a photonic alloy composed of PEC and magnetized YIG rods.  (b)-(d) Field distributions $|E_z|$ when the doping concentration $x=0, 0.5, 1$ at 11.60 GHz, respectively. The white dots indicate the position of PEC rods, and the blue stars indicate the position of line sources. The top, bottom and left sides of the sample are covered with PEC, and the right side is covered with a wave-absorbing material.}\label{Fig1}
\end{figure*}	

In this paper, we answer this question in an affirmative way by presenting the creation of a topological photonic alloy by mixing Perfect Electric Conductor (PEC, $A$ component) and magnetized YIG ($B$ component) rods in a square lattice, as shown in Fig. \ref{Fig1}(a). It is found that the electromagnetic wave at frequency 11.60 GHz propagate in the bulk region of the PEC and magnetized YIG photonic crystal, as shown in Figs. \ref{Fig1}(b) and (d). The photonic alloy system with random doping of the two components ($x=0.5$) can form a chiral edge state (CES) at the same frequency, as shown in Fig. \ref{Fig1}(c). Here the doping concentration $x$ is defined as $x = N_B/(N_A+N_B)$ with $N_A$ and $N_B$ representing the number of $A$-type (PEC) and $B$-type (magnetized YIG) rods. In order to explain this phenomenon, we firstly analyze the band structures of square lattices under two extreme cases, a lattice composed entirely of PEC or magnetized YIG rods. We discovered that the bandgaps of these two lattices do not overlap in frequency at 11.60 GHz. The PEC lattice exhibits a topologically trivial bandgap in the frequency range of 11.83 GHz to 12.19 GHz, while the magnetized YIG lattice possesses a topologically non-trivial bandgap in the frequency range of 10.87 GHz to 11.53 GHz. There is a shift between these two bandgaps such that the bandgaps lack an overlapping frequency region, which is analogous to the broken-gap type band alignment in semiconductor heterostructures \cite{yoshitake2021work,bandoffset}. Additionally, a phenomenon analogous to band inversion at the $M$ point is observed for the bandgaps of the PEC and magnetized YIG lattices. Next, the transmission spectrum at different doping concentrations was calculated. The frequency range of the photonic alloy transmission gap extends gradually from the frequency of the PEC bandgap to the bandgap of the magnetized YIG rods, exhibiting an overall downward trend with the increasing doping concentration. Additionally, a gradual closing and reopening of the transmission gap in this process was observed. The topological transition process is validated using the scattering method. Finally, the emergence of edge states during this process and their variation with concentration $x$ from a momentum space perspective is studied.

\section{Model and Results}

\begin{figure}
	\centering
	\includegraphics[width=0.5\textwidth]{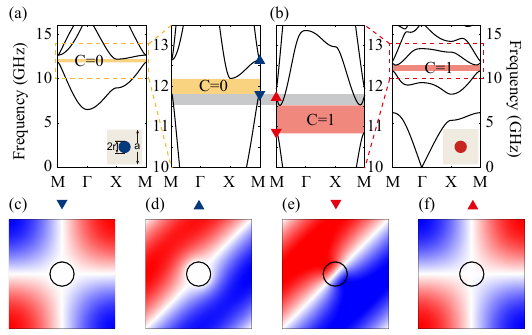}
	\caption{(a)-(b) The TM bulk band structures of PEC/magnetized YIG photonic crystal, respectively. The lower inset illustrates the lattice geometry. (c)-(d) The distributions of the eigenmode field $E_z$ corresponding to the first (blue downward triangle) and second (blue upward triangle) photonic band at the $M$ point of the PEC photonic crystal. (e)-(f) The distributions of the eigenmode field $E_z$ corresponding to the second (red downward triangle) and third (red upward triangle) photonic band at the $M$ point of the magnetized YIG photonic crystal.}\label{Fig2}
\end{figure}	

Magnetized YIG material exhibits a magnetic response only within the microwave frequency range. Metals such as gold, silver, and copper are considered as PEC within this same frequency range \cite{Wang9,Skirlo45,Skirlo46}. The relative permittivity of the YIG rods is 14.5. A static magnetic field, generated by a permanent magnet, is applied perpendicularly to each YIG rod. The saturation magnetization $4\pi M_s=1850$G , and the gyromagnetic resonance loss width  $\Delta H=50$Oe. The magnetic permeability tensor of the magnetized YIG rods assumes the following form:
\begin{equation}
	\mu=\left(\begin{matrix}\mu_1&{i\mu}_2&0\\{-i\mu}_2&\mu_1&0\\0&0&\mu_0\\\end{matrix}\right)
\end{equation}
where $\mu_1=\mu_0\left(1+\frac{\omega_0\omega_m}{\omega_0^2-\omega^2}\right)$, $\mu_2=\mu_0\frac{\omega\omega_m}{\omega_0^2-\omega^2}$,  $\mu_0$ is the permeability of vacuum. The resonance frequency $\omega_0=\gamma H_0$, the characteristic frequency $\omega_{m}=4\pi\gamma M_s$, the gyromagnetic ratio $\gamma=2.8$ MHz/Oe and the external magnetic field $H_0 = 800$
Oe along the $z$ direction. Loss is included by taking $\omega_0\rightarrow\omega_0+i\frac{\gamma\Delta H}{2}$. 
The operating frequency near the center of the magnetized YIG gap 11.15 GHz. 
The material parameters regarding YIG such as saturation magnetization, gyromagnetic resonance loss width, and applied magnetic field are the parameter values used in the experiment \cite{TPA}. By using these parameters, the numerical simulations can be directly compared with the experiment results.
We employ finite-element methods to numerically solve Maxwell's equations with various boundary conditions and with or without an excited source, using COMSOL Multiphysics. After computing the electromagnetic field or their corresponding eigenfrequencies via COMSOL, we numerically calculate the band structure, transmission spectrum, reflection phases, and Fourier transformation of the electromagnetic field into momentum space in the following.

\begin{figure}
	\centering
	\includegraphics[width=0.5\textwidth]{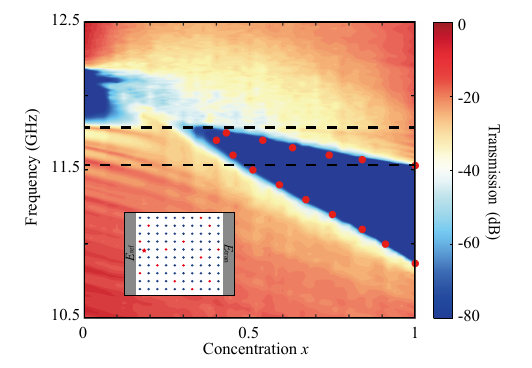}
	\caption{The simulated transmission spectrum versus $x$ is averaged over 20 disorder samples.  The red circles represent the topological gap calculated from the reflection phase winding. Insets: schematic of numerically calculated transmission in topological photonic alloy systems. The upper and lower boundaries are continuous, the left and right boundaries are absorbing, and line source location is marked by a red star.}\label{Fig3}
\end{figure}	

\begin{figure}
	\centering
	\includegraphics[width=0.45\textwidth]{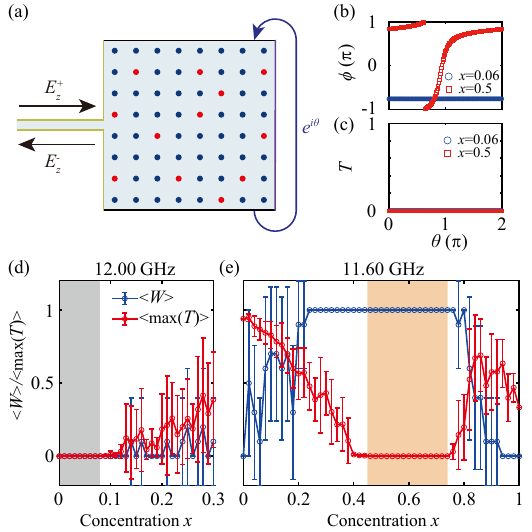}
	\caption{(a) Schematic for retrieving the topological signature of CES from the reflection phase connected to a square photonic alloy with linear size $W$ with a twisted boundary condition  $\Psi(y=L)=\Psi(y=0)e^{i\theta}$ imposed to the vertical boundaries. The left side of the photonic alloys is connected with an air lead bounded by PMC (yellow), and the right (purple) boundary is set as scattering boundary condition. (b)-(c) The blue circles (red squares) represent the variation of reflection phase $\phi$ and transmittance $T$ with respect to the change of $\theta$ at 12.00 GHz when $x=0.06$ (at 11.60 GHz when $x=0.5$), respectively. The winding number of the reflection phase and the maximum transmittance $\langle {\rm max}(T)\rangle$ vary with the doping concentration $x$ at 12.00 GHz (d) and 11.60 GHz (e), where each data point represents the average of ten disorder samples.}\label{Fig4}
\end{figure}	

The transverse magnetic (TM) bulk band structure of the square lattice composed of PEC rods is shown in Fig. \ref{Fig2}(a), where the radius $r=2$mm, and lattice constant $a=18$mm. Contrary to photonic crystals made of conventional dielectric materials, there are no eigenstates near the low frequency at the $\Gamma$ point \cite{metal_PC}. The entire system maintains time-reversal symmetry, so the band gap in the frequency range of [11.83, 12.19] GHz is topologically trivial with a Chern number of 0, situated between the first and second bands. The eigenstates of the first and second bands at the $M$ point are $d$-like and $p$-like bands, as shown in Figs. \ref{Fig2}(c) and (d) marked by blue downward and upward triangles in Fig. \ref{Fig2}(a), respectively.
The band structure of the square photonic crystal made of magnetized YIG material is depicted in Fig. \ref{Fig2}(b), with the same radius and lattice constant as the PEC photonic crystal. There is a topological band gap in the frequency range of [10.87, 11.53] GHz with a Chern number of 1, situated between the second and third bands. The eigenstates of the second and third bands at the $M$ point are $p$-like and $d$-like bands, as shown in Figs. \ref{Fig2}(e) and (f) marked by red downward and upward triangles in Fig. \ref{Fig2}(b), respectively. The field distribution of the eigenstates at the $M$ point indicates a phenomenon similar to band inversion in these two band gaps. From the band structures, the lack of overlap between these two bandgaps is analogous to the broken gap type band alignment in semiconductor heterostructures \cite{yoshitake2021work,bandoffset}. In the following, we will investigate the band alignment effect in the topological photonic alloy.

Previous studies have achieved band inversion in the same crystal by gradually adjusting parameters like rod radius and position, often resulting in a topological phase transition \cite{Skirlo45,Hu2015,Liu2020}. However, the importance of band inversion between different crystals has not been extensively explored.
The primary challenge in investigating this issue lies in the difficulty of finding a continuous and adjustable parameter to transform between the square lattice composed of PEC rods and the one composed of magnetized YIG rods. The emergence of topological photonic alloy systems provides a new platform to the transformation between two completely different crystals. Specifically, continuously dopes one type of rods into the supercell of another type, as illustrated in Fig. \ref{Fig1}(a), with red circles representing magnetized YIG rods and blue circles representing PEC rods. In the absence of doping, it corresponds to a PEC photonic crystal, while with complete doping, it corresponds to a magnetized YIG photonic crystal. The gradual transition from one type of photonic crystal to another is achieved by adjusting the doping concentration $x$.

Now, we demonstrate the corresponding bulk transmission of alloy systems composed of two distinctly different crystals (PEC and magnetized YIG lattices) with the increase in doping concentration $x$, as shown in Fig. \ref{Fig3}. The square sample size $L$ used is $30a \times 30a$.
We wish to clarify that $x$ changes discretely in the finite system with a discrete interval $1/(N_A+N_B )$. In the numerical results presented in Fig. \ref{Fig3}, the doping concentration $x$ changes with a step size of 0.02 due to the high computational cost.
To measure bulk transmission, we connect the top and bottom of the photonic alloy with continuous boundaries, using absorbing boundaries on the left and right sides, a line source is placed near the left boundary, as shown in the inset of Fig. \ref{Fig3}. 
And then we integrate the Poynting vectors at the left and right boundaries to obtain the time-averaged energies. $E_{\rm tran}$ represents the energy that passes through the photonic alloy. $E_{\rm ref}$ is the energy that is directly reflected by the photonic alloy. The total energy flowing out from both the left and right boundaries is defined as $E_{\rm tot} = E_{\rm ref}+E_{\rm tran}$.
The bulk transmission is obtained through $\langle T\rangle=\langle {\rm 20  log}_{10}(E_{\rm tran}/E_{\rm tot})\rangle$ \cite{Skirlo46}. From the calculation of the transmission spectrum, the transmission gap of the topological photonic alloy converges to the bandgap of the PEC lattice as $x$ tends to 0, while it converges to the bandgap of the magnetized YIG lattice as $x$ tends to 1. During the process of increasing $x$, there is a phenomenon of gap closure followed by reopening, indicating that these two gaps may possess different topological properties. In the frequency intervals region (between the black dash lines), the photonic alloy is able to realize the CES by mixing at a certain concentration when the components are supporting the bulk transport as shown in Fig. \ref{Fig1}(b)-(d).

Next, we characterize the topological properties of photonic alloy systems by scattering method \cite{scatter2010,scatter2011,scatter2012,Zhe26,Cui2022}.
Specifically, we imposed a twisted boundary condition $\Psi(y=L)=\Psi(y=0)e^{i\theta}$ on the vertical boundaries \cite{Cui2022}. The left side of the photonic alloys is connected to an air lead, bounded by Perfect Magnetic Conductor (PMC). The right boundary is set as an absorbing boundary condition, as illustrated in Fig. \ref{Fig4}(a).
Figure \ref{Fig4}(b) shows the change in reflection phase with the twist angle $\theta$. The blue circle signifies the reflection phase when $x=0.06$ at 12.00 GHz, consistently oscillating around a specific value. The red square indicates the reflection phase when $x=0.5$ at 11.60 GHz. 
As $\theta$ varies from 0 to $2\pi$, the reflection phase $\phi(\theta)$ changes cumulatively by $2\pi$, so that the winding number of reflection phase $W=1$ (More details can be found in Appendix C).
Additionally, the corresponding transmittance $T$ for both cases is zero, as depicted in Fig. \ref{Fig4}(c). These suggest that the topological invariants of the photonic alloy system differ under these two parameter sets.

By taking a closer look, we calculated the winding number of the reflection phase and the maximum transmittance at 12.00 GHz as a function of doping concentration $x$, as depicted in Fig. \ref{Fig4}(d). In the range of $x\in[0,0.08]$, the maximum transmittance for ten samples is nearly zero, with minimal fluctuations. Here, the reflection phase remains constant and does not complete a $2\pi$ rotation, indicating a topologically trivial gap. However, when $x>0.08$, a significant transmittance emerges ($\langle {\rm max} (T) \rangle>0$), implying that photons can pass through the bulk. We also calculated the winding number of the reflection phase at 11.60 GHz as a function of doping concentration $x$, as displayed in Fig. \ref{Fig4}(e). For $x\in [0.45,0.74]$, the transmittance is nearly zero. In this range, the reflection phase for corresponding samples accumulates $2\pi$, which indicates a topologically nontrivial gap. Corresponding to real space, there is a CES between the photonic alloy and PEC boundaries, as shown in Fig. \ref{Fig1}(c). When $x<0.45$ or $x>0.74$, photons can pass through the bulk. Additionally, we used scattering methods to calculate the boundaries of the topological gaps, which are marked with red circles in Fig. \ref{Fig3}.
Notably, the transition from a topologically trivial gap to a non-trivial one in this process differs from the topological Anderson insulator in disordered photonic crystals from a symmetry perspective \cite{Liu35}. The formation of topological Anderson insulators in disordered photonic crystals is driven by the competition between time-inversion and spatial inversion symmetry. While our model maintains spatial inversion symmetry on average over multiple samples \cite{disorderavg2012,disorderavg2013,disorderavg2021,disorderavg2022}.

\begin{figure*}
	\centering
	\includegraphics[width=0.95\textwidth]{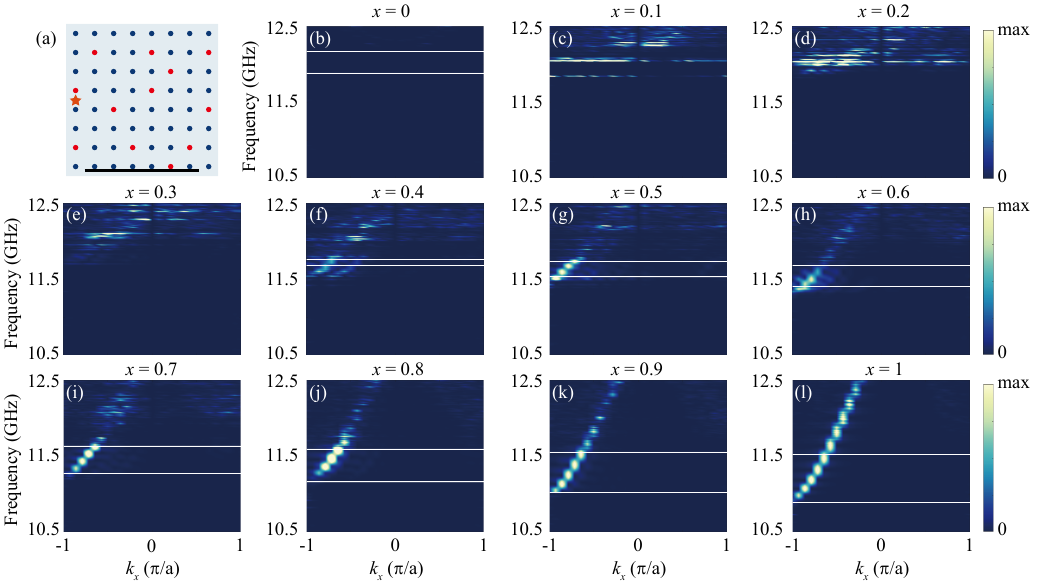}
	\caption{(a) Schematic diagram for calculating the edge state in momentum space. The orange star marks the position of the line source. The black line indicates the location of the extracted electric field data.  (b)-(l) The edge state in momentum space for $x = 0, 0.1, 0.2, 0.3, 0.4, 0.5, 0.6, 0.7, 0.8, 0.9, 1$, respectively. The solid white solid lines indicate the average transmission gaps across 20 samples.}\label{Fig5}
\end{figure*}	

Finally, by applying Fourier transformation to real-space electromagnetic fields, the gradual appearance of the CES is identified in momentum space. Figure \ref{Fig5}(a) presents a schematic of the calculated edge states in momentum space, with boundary conditions identical to those in Figs. \ref{Fig1}(b)-(d). The orange star denotes the line source's position, and the black solid line along the bottom boundary indicates the position for extracting the edge states' electric field. As there is no electric field distribution in the PEC rod, we chose a location 6 mm from the PEC boundary (3 mm from the center of lowermost rods) and a straight line 26$a$ in length.
By Fourier transforming the extracted electric field data, we can ascertain the edge states in momentum space. Initially, with $x=0$, almost no band is present in the bulk transmission gap (between the two white solid lines), as depicted in Fig. \ref{Fig5}(b). This lack of band occurs because edge states do not form in topologically trivial crystals, as indicated in Fig. \ref{Fig1}(b). However, at $x=1$, a continuous band emerges in the band gap region (between the two white solid lines), as depicted in Fig. \ref{Fig5}(l).
Beyond the upper boundary of the gap, the edge state within the bulk band is still observable, corresponding to the real space as shown in Fig. \ref{Fig1}(d). As $x$ increases to 0.4, topological edge states appear in the gap region. As $x$ continues to increase, the gap broadens, making the topological edge states more evident, which is a manifestation of the band alignment effect in momentum space. Moreover, in the range $0.4 \leq x \leq 0.9$, topological edge states in the bulk band are subdued due to the crystal's disordered arrangement, as depicted in Figs. \ref{Fig5}(f)-(k). The curves of topological edge states in momentum space beyond the upper gap become dark, even chaotic, yet the curves in the gap region manage to retain clear band-like structures.

\section{CONCLUSIONS AND OUTLOOK}
In summary, we have created a topological photonic alloy by randomly doping PEC and magnetized YIG rods in a square lattice. It is worth noting that the edge states of the topological photonics alloy can appear at the frequency range where both constituent components are in their bulk states. We analyze the PEC lattice and magnetized YIG lattice band gaps appearing in a broken gap type of band alignment and there is a band inversion-like phenomenon between these two lattices at the $M$ point. Subsequently, the transmission spectrum, calculated at different doping concentrations, shows the photonic alloy transmission gap shifting from the PEC bandgap to the magnetized YIG bandgap, gradually closing and reopening the gap. Furthermore, the emergence of robust one-way edge states in the topological gap where $x$ ranges from 0.4 to 1 was verified, and the topological invariant was calculated using scattering methods. Finally, by Fourier transforming the real-space fields, the edge states in momentum space were obtained, confirming the gradual appearance of edge states with increasing doping concentration $x$ in these photonic alloys. The proposed experimental setup is simple, utilizing common materials, and it is hoped that it can be implemented in the near future on microwave photonic experimental platforms. The noteworthy point is that the alloy doping method is not limited to magnetized materials and can be extended to ordinary dielectric materials, allowing topological photonic alloys to be generalized to the optical frequency range. Looking ahead, combining deep learning for the reverse design of photonic crystals with different band structures \cite{deep1,deep2,deep3,deep4} and leveraging alloy doping holds the promise of achieving even more diverse topological effects. This lays the foundation for the flexible application of topological physics in photonic devices.

This work is supported by the National Natural Science Foundation of China Grant No. 12074230, 12174231, the Fund for Shanxi ``1331 Project", Fundamental Research Program of Shanxi Province through 202103021222001. This research was partially conducted using the High Performance Computer of Shanxi University.

\bigskip
\noindent{$^{*)}$mudiwang@ust.hk}\\
\noindent{$^{\dagger)}$chenjun@sxu.edu.cn}\\
\noindent{$^{\ddagger)}$zhanglei@sxu.edu.cn}

\appendix

\section{Numerical calculation of photonic band structures}
We choose the primitive cell in the PEC/magnetized YIG photonic crystal, as shown in the insets of Figs. \ref{Fig2} (a, b). We connect the top and bottom and left and right boundaries of the primitive cell with periodic boundaries. 
The $x$- and $y$-direction Bloch wave vectors $k_x$ and $k_y$  are set so that a path in the momentum space from $\Gamma$ ($k_x=0$,  $k_y=0$) to $X$ ($k_x=\pi/a$,  $k_y=0$) and then to $M$ ($k_x=\pi/a$ ,  $k_y=\pi/a$) and finally back to the $\Gamma$ point in the first Brillouin zone. Finally, the variation of the eigenfrequencies with $k$-point paths is plotted to obtain the band structure shown in Figs. \ref{Fig2}(a, b).

\section{The Chern number for periodic photonic crystal}
In the formulation of Maxwell’s equation, the periodic part of the eigenfunctions of the electric field $\boldsymbol{u}_n (\boldsymbol{k},\boldsymbol{r})= \langle \boldsymbol{r} | \boldsymbol{u}_n (\boldsymbol{k}) \rangle$ is the 3-component vector of complex electromagnetic fields \cite{Haldane8}, as shown in below
\begin{equation}
	\boldsymbol{u}_n(\boldsymbol{k},\boldsymbol{r})=\boldsymbol{E}_n(\boldsymbol{k},\boldsymbol{r}) e^{-i\boldsymbol{k} \cdot \boldsymbol{r}}.
\end{equation}
For TM modes, the magnetic field is confined to the $x$-$y$ plane. Therefore, the non-zero components of the magnetic field are $H_x(r)$, $H_y(r)$. Meanwhile, the electric field becomes a scalar function, being $E_z (\boldsymbol{r})$ the only nonzero vector component. The electromagnetic field are numerically calculated by COMSOL. 

The Chern number of the $n$th photonic band for 2D systems is defined as
\begin{equation}\label{chern_number}
	C_n=\frac1{2\pi}\oint_{BZ}\Omega_n^z(\boldsymbol{k})d^2\boldsymbol{k} ,
\end{equation}
\begin{equation}\label{berry_curvature}
	\boldsymbol{\Omega}_n(\boldsymbol{k})=i\nabla_{\boldsymbol{k}}\times\langle\boldsymbol{u}_n(\boldsymbol{k},\boldsymbol{r})|\nabla_{\boldsymbol{k}}|\boldsymbol{u}_n(\boldsymbol{k},\boldsymbol{r})\rangle,
\end{equation}
where $\Omega_n^z$ is the $z$ component of Berry curvature $\boldsymbol{\Omega}_n (\boldsymbol{k})$, the $k$-space integral in Eq. (\ref{chern_number}) is performed over the first Brillouin zone with [$k_x$, $k_y$], and the Berry curvature is defined as Eq. (\ref{berry_curvature}) \cite{Lu6}. The gap Chern number ($C_{\rm{gap}}=\sum_n C_n $) is to sum the Chern numbers of all bands below the band gap \cite{Skirlo45}.

\section{The winding number of the reflection phase of non-periodic photonic systems}
In this section, we introduce the winding number of the reflection phase can be used to characterize the topology of non-periodic optical systems.
In the scattering process of an incident field $E_z^+(\theta,\omega)$ impinging upon the photonic alloy, a reflection matrix $R$ relates the reflected wave to the incident wave as  $E_z^{-}(\theta, \omega)=R E_z^{+}(\theta, \omega)$, as shown in Fig. \ref{Fig4}(a) of the main text. The reflection matrix $R$ is a function of frequency and twist angle $\theta$ of the boundary condition. Inside the gap of the photonic alloy, $R=e^{i\phi} \in \rm{U(1)}$. When we adiabatically vary $\theta$ over a period (from $\theta$ to $2\pi$), the first homotopy group of $\pi_1(\rm{U(1)})=\mathbb{Z}$ characterizes the topological classification of the evolutions of reflection $R$. Denoting by the integer winding number of the reflection phase $\phi(\theta)=arg(R(\theta))$ is 
\begin{equation}
	W=\frac{-i}{2 \pi} \int_0^{2 \pi} \frac{\partial \ln R}{\partial \theta} d \theta=\frac{1}{2 \pi} \int_0^{2 \pi} \frac{\partial \phi}{\partial \theta} d \theta \in \mathbb{Z}
\end{equation}
Physically, the application of twisted boundary conditions can be viewed as an adiabatically changing gauge flux (black) $\theta$ threading the hollow of a rolled-up photonic alloy. The winding of the reflection phase method can be effectively viewed as a “topological quantum pump” \cite{pump1,pump2}. The results of reflection phases are presented in the Fig. \ref{Fig4}. Given a twisting boundary angle $\theta$, an electric field $E_z$ is incident from the left port as shown in Fig. \ref{Fig4}(a). The reflection phase is finally found by examining the scattering matrix. As $\theta$ changes from 0 to $2\pi$, the reflection phase $\phi(\theta)$ changes cumulatively by $2\pi$ in the gap, similar to the Wilson loop, indicating that the system is topologically non-trivial. 

\bibliography{myref}
\end{document}